\documentstyle[anta, twoside]{article}

\input{psfig.sty}

\def\Journal#1#2#3#4{(#1) {#2} {\bf #3}, #4}

\def\AJ{\em Astron.~J.}
\def\ApJ{\em Astrophys.~J.}
\def\ApJL{\em Astrophys.~J., Lett.}

\def\AaAp{\em Astron. Astrophys.}

\def\MNRAS{\em Mon. Not. R.~Astron. Soc.}
\def\Nat{\em Nature\/}

\begin{document}

\markboth{Fedor V.Prigara}{The Wavelength Dependence of
Polarization in Active Galactic Nuclei}

\thispagestyle{plain}

\title{The Wavelength Dependence of Polarization in Active Galactic Nuclei}

\author{Fedor V.Prigara}

\address{Institute of Microelectronics and Informatics, Russian Academy of Sciences,
21 Universitetskaya, 150007 Yaroslavl, Russia}

\maketitle

\abstract{Using the gaseous disk model and the condition for
emission based on a stimulated origin of thermal radio emission
(Prigara 2003), we received the wavelength dependence of
polarization for radio emission from active galactic nuclei
consistent with available observational data. The strength of
magnetic field required to produce an observable degree of
polarization in active galactic nuclei is discussed.}

\section{Introduction}

The degree of polarization for thermal radiation in a magnetic field was
obtained by J.C.Kemp in1970 (for review see Lang 1974). It is given by the
formula

\begin{equation}
\label{eq1}
p = eB/2mc\omega
\end{equation}

\noindent
where \textit{e} and \textit{m} are the charge and mass of electron
respectively, \textit{c} is the speed of light, \textit{B} is the strength
of magnetic field, and \textit{w} is the circular frequency of radiation.

According to equation (1), the wavelength dependence of a degree
of polarization is determined by the magnetic field profile
\textit{B(r)}, where \textit{r} is the distance from a central
energy source, and by the relationship between the frequency of
radiation and the radius \textit{r}. The last relationship is
given by the condition for emission (Prigara 2003). To introduce
the magnetic field profile we consider the gaseous disk model for
radio source.

\section{The gaseous disk model}

It was shown recently (Prigara 2003) that thermal radio emission
has a stimulated character. According to this conception thermal
radio emission of non-uniform gas is produced by an ensemble of
individual emitters. Each of these emitters is a molecular
resonator the size of which has an order of magnitude of mean free
path \textit{l} of photons

\begin{equation}
\label{eq2}
l = \frac{{1}}{{n\sigma} }
\end{equation}

\noindent
where \textit{n} is the number density of particles and $\sigma $ is the
absorption cross-section.

The emission of each molecular resonator is coherent, with the wavelength

\begin{equation}
\label{eq3}
\lambda = l,
\end{equation}

\noindent
and thermal radio emission of gaseous layer is incoherent sum of radiation
produced by individual emitters.

The condition (3) implies that the radiation with the wavelength
$\lambda $ is produced by the gaseous layer with the definite
number density of particles \textit{n} .

The condition (3) is consistent with the experimental results by
Looney and Brown on the excitation of plasma waves by electron
beam (see Chen 1987, Alexeev 2003). The wavelength of standing
wave with the Langmuir frequency of oscillations depends on the
density as predicted by equation (2). The discreet spectrum of
oscillations is produced by the non-uniformity of plasma and the
reajustment of the wavelength to the length of resonator. From the
results of experiment by Looney and Brown the absorption
cross-section for plasma can be evaluated.

The product of the wavelength by density is weakly increasing with the
increase of density. This may imply the weak dependence of the size of
elementary resonator in terms of the wavelength upon the density or,
equivalently, wavelength.

In the gaseous disk model, describing radio emitting gas nebulae
(Prigara 2003), the number density of particles decreases
reciprocally with respect to the distance \textit{r} from the
energy center

\begin{equation}
\label{eq4}
n \propto r^{ - 1}.
\end{equation}

Together with the condition for emission (3) the last equation
leads to the wavelength dependence of radio source size:

\begin{equation}
\label{eq5}
r_{\lambda}  \propto \lambda .
\end{equation}

The relation (5) is indeed observed for sufficiently extended
radio sources. For example, the size of radio core of galaxy M31
is 3.5 arcmin at the frequency 408 MHz and 1 arcmin at the
frequency 1407 MHz (Sharov 1982).

\section{The wavelength dependence of radio source size}

In the case of compact radio sources instead of the relationship
(4) the relationship

\begin{equation}
\label{eq6}
r_{\lambda}  \propto \lambda ^{2}
\end{equation}

\noindent is observed (Lo et al. 1993, Lo 1982). This relationship
may be explained by the effect of a gravitational field on the
motion of gas which changes the equation (3) for the equation
\begin{equation}
\label{eq7}
 n \propto r^{-1/2}.
\end{equation}
The mass conservation in an outflow or inflow of gas gives
\textit{nvr=const,} where \textit{v} is the velocity of flow. In the
gravitational field of a central energy source the energy conservation gives

\begin{equation}
\label{eq8}
v = \left( {v_{0}^{2} + c^{2}r_{s} /r} \right)^{1/2}
\end{equation}

\noindent where $r_{s} $ is the Schwarzschild radius. Therefore,
at small values of the radius the equation (7) is valid, whereas
at the larger radii we obtain the equation (4).
 The relationship
between linear size and turnover frequency in gigahertz-peaked
spectrum sources and steep-spectrum sources (Nagar et al. 2002) is
a consequence of the wavelength dependence of radio source size.
The turnover frequency is determined by the equation $r_{\nu} =
R$, where R is the radius of a gaseous disk. The same equation
determines a turnover frequency for planetary nebulae (Pottasch
1984; Siodmiak \& Tylenda 2001).

To summarize, extended radio sources are characterized by the
relation (5), and compact radio sources obey the relation (6).

\section{The wavelength dependence of polarization}

At first, consider extended radio sources for which the equation
(5) is valid. Assuming the magnetic field profile $B \propto r^{ -
2}_{^{}} $, which is characteristic for the field of
two-dimensional magnetic dipole, from equation (1) we obtain

\begin{equation}
\label{eq9}
p \propto B\left( {r_{\lambda} }  \right)\lambda
\propto \lambda ^{ - 1}.
\end{equation}

This relationship is qualitatively consistent with observational
data. The degree of linear and circular polarization in active
galactic nuclei typically decreases with the increase of
wavelength (Bower et al. 1999; Bower et al. 2001). In the case of
Cen A the law (9) is valid in the decimeter range.

For compact radio sources the law (6) is valid, and the magnetic
field profile is also changed to $B \propto r^{ - 3/2}$ (Prigara
2003). Now the equation (1) gives

\begin{equation}
\label{eq10}
 p \propto \lambda^{-2}.
\end{equation}

The degree of linear polarization in 3C 84 is \textit{p}=0.03\% at
the frequency 4.8 GHz and \textit{p}>0.2\% at the frequency 14.5
GHz (Bower et al. 1999), in agreement with the equation (10). The
degree of linear polarization in Cen A in centimeter range also
follows the equation (10).

It is known that the synchrotron theory predicts a change in the
polarization position angle across the spectral peak of gigahertz-peaked
source. No such a change was found for the six gigahertz-peaked sources
(Mutoh et al. 2002).

\section{The strength of magnetic field}

According to equation (1) the field \textit{B}=20 G is required to
produce the degree of polarization \textit{p}=0.2\% at the
frequency 14.5 GHz, as in the case of 3C 84 (Bower et al. 1999).
If we assume, for compact radio sources, that $r_{\lambda} = r_{0}
$ at the wavelength $\lambda $=1 mm, where $r_{0} $ is the radius
of the photosphere of a central energy source (e.g., the
Schwarzschild radius), then $B_{0} = B_{\lambda}  \left( {\lambda
/\lambda _{0}}  \right)^{3} = 2 \times 10^{5}G$. Zakharov et al.
(2003) have obtained an estimate $B < 10^{10} - 10^{11}G$ for the
Seyfert galaxy MCG-6-30-15.

The existence of regular magnetic fields in active galactic nuclei is
supported by the geometry of radio jets following the lines of a magnetic
field. In many cases jets have a significant curvature, sometimes up to 90
degrees or more (Kellermann et al. 1998). The stable handedness of circular
polarization in Sgr A* and M81* requires stable global magnetic field
components (Beckert \& Falcke 2002, Bower, Falcke \& Mellon 2002). This
suggests the existence of regular magnetic fields in accretion flows too.

One of the possible explanations for line widths in active galactic nuclei
is a magnetic broadening. A magnetic broadening of spectral lines is
produced by the Zeeman splitting in a non-uniform magnetic field changing
its strength in the emitting region. The magnetic broadening in AGN is
supported by broader $H_{\alpha}  $ lines in polarized emission than in
total emission (Nagar et al. 2002; Hes, Barthel \& Fosbury 1993). However,
the magnetic broadening of broad emission lines in AGNs encounters some
difficulties. First, a change in the continuum flux produces an earlier
response in the red wing of the line than in the blue wing (e.g., Konigl
2003). Second, some AGNs show double-peaked broad emission lines (Strateva
et al. 2003). Both these features are indicative of a convection in the disk
corona. In particular, the double-peaked emission lines are observed in
planetary nebulae (Pottasch 1984). Nevertheless, a magnetic broadening can
be applied to the narrow emission lines in AGNs. It may also participate in
the width of each component of a double-peaked emission line.

A magnetic broadening may be a solution to the well-known problem of line
broadening in A, B and O stars (Bohm-Vitense 1980). The Ca II lines in B
stars usually assumed to be interstellar alternatively may be interpreted as
a result of a non-uniform shift of spectral lines (Ebbets 1980) together
with the Zeeman splitting. The latter produces the several (6 to 10)
components of Ca II lines.

The phase diagram for the matter in ultrastrong magnetic fields is virtually
unknown. The phase transitions in sufficiently strong magnetic fields may be
responsible both for hard gamma-ray flares (Kellermann et al. 1998) and
cosmic rays of ultrahigh energies. The origin of the latter is not
elucidated up to date.

The effect of the strong magnetic field on the matter is produced
by the magnetic pressure. In magneto-hydrodynamics it is assumed
that the magnetic pressure is proportional to the $B^{2}$.
One-fluid magneto-hydrodynamics is, however, poorly justified
(Kadomtsev 1988). Alternatively, the magnetic pressure in a hot
plasma may be suggested to be proportional to the strength of a
magnetic field as follows

\begin{equation}
\label{eq11}
p_{B} = eBv{}_{0}/\sigma c,
\end{equation}

\noindent where $\sigma $ is the collisional cross-section in the
high-temperature limit (when $e^{2}/kT<h^{2}/Zme^{2}$), and $v_{0}
$ is a constant.

In the case of flat-spectrum active galactic nuclei (Bower \&
Backer 1998; Nagar, Wilson \& Falcke 2001; Ulvestad \& Ho 2001)
the density, temperature, and pressure profiles have a form $n
\propto r^{ - 1/2},T \propto r^{ - 1},P = nkT \propto r^{ - 3/2}$.
Since for compact radio sources $B \propto r^{ - 3/2}$, from
equation (11) we obtain $P \propto P_{B} ^{}$. The latter
relationship confirms equation (11). Here the density, temperature
and pressure profiles are the same as those in the
convection-dominated accretion flow (CDAF) models (Nagar et al.
2001). The magnetic field profile is different since the
expression for the magnetic pressure is changed. The ratio of
magnetic to gas pressure will be fixed similar to the CDAF models.

\section{Conclusions}

The maser theory of thermal radio emission (Prigara 2003) can
explain the wavelength dependence of polarization in active
galactic nuclei, if we only assume the existence of sufficiently
strong magnetic fields in AGN. The other indications of strong
magnetic fields in AGN are a magnetic broadening of spectral lines
and possibly the hard gamma-ray flares.

\section*{References}\noindent

\references

Alexeev B.V. \Journal{2003}{\em Usp. Fiz. Nauk}{173}{145}.

Beckert T., Falcke H. \Journal{2002}{\AaAp}{388}{1106}.

Bohm-Vitense E. (1980), in {\em Stellar Turbulence, IAU Colloq.
51}, eds. D.F.Gray, J.L.Linsky, Springer.

Bower G.C., Backer D.C. \Journal{1998}{\ApJL}{507}{L117}.

Bower G.C., Falcke H., Mellon R.R.
\Journal{2002}{\ApJL}{578}{L103}.

Bower G.C., Wright M.C.H., Backer D.C., Falcke H.
\Journal{1999}{\ApJ}{527}{851}.

Bower G.C., Wright M.C.H., Falcke H., Backer D.C.
\Journal{2001}{\ApJL}{555}{L103}.

Chen F.F. (1984), {\em Introduction to Plasma Physics and
Controlled Fusion, Vol.1: Plasma Physics}, Plenum Press.

Ebbets D. (1980) in {\em Stellar Turbulence, IAU Colloq. 51}, eds.
D.F.Gray, J.L.Linsky, Springer.

Hes R., Barthel P.D., Fosbury R.A.E.
\Journal{1993}{\Nat}{361}{326}.

Kadomtsev B.B. (1988), {\em Cooperative phenomena in plasmas},
Nauka.

Kellermann K.I., Vermeulen R.C., Zensus J.A., Cohen M.H.
\Journal{1998}{\AJ}{115}{1295}.

Konigl A. (2003) in {\em Active Galactic Nuclei: from Central
Engine to Host Galaxy, ASP Conf. Series 290}, eds. S.Collin,
F.Combes, I.Shlosman, ASP.

Lang K.R. (1974), {\em Astrophysical Formulae}, Springer.

Lo K.Y. (1982) in {\em AIP Proc. 83: The Galactic Center}, eds.
G.Riegler, R.Blandford, AIP.

Lo K.Y., Backer D.C., Kellermann K.I., Reid M., Zhao J.H., Goss
M.H., Moran J.M. \Journal{1993}{\Nat}{361}{38}.

Mutoh M., Inoue M., Kameno S., Asada K., Fujisawa K., Uchida Y.
(2002), {\em astro-ph/0201144}.

Nagar N.M., Wilson A.S., Falcke H.
\Journal{2001}{\ApJL}{559}{L87}.

Nagar N.M., Wilson A.S., Falcke H., Ulvestad J.S., Mundell C.G.
(2002) in {\em Issues in Unification of AGNs, ASP Conf. Series
258}, eds. R.Maiolino, A.Marconi, N.Nagar, ASP.

Pottasch S.R. (1984), {\em Planetary Nebulae}, D.Reidel Reinhold
Company.

Prigara F.V. (2003), {\em Astron. Nachr.}, Vol. 324, No. S1, {\em
Supplement: Proceedings of the Galactic Center Workshop 2002 - The
central 300 parsecs of the Milky Way}, eds. A.Cotera, H.Falcke,
T.R.Geballe, S.Markoff.

Sharov A.S. (1982), {\em The Andromeda Nebula}, Nauka.

Siodmiak N., Tylenda R. \Journal{2001}{\AaAp}{373}{1032}.

Strateva I., Strauss M.A., Hao L. (2003) in {\em Carnegie
Observatories Astrophysics Series, Vol.1: Coevolution of Black
Holes and Galaxies}, ed. L.C.Ho, Carnegie Observatories.

Tadhunter C., Wills K., Morganti R., Oosterloo T., Dickson R.
\Journal{2001}{\MNRAS}{327}{227}.

Ulvestad J.S., Ho L.C. \Journal{2001}{\ApJL}{562}{L133}.

Zakharov A.F., Kardashev N.S., Lukash V.N., Repin S.V.
\Journal{2003}{\MNRAS}{342}{1325}.

\end{document}